\documentclass[a4paper,fleqn,usenatbib]{mnras}

\usepackage[T1]{fontenc}
\usepackage{ae,aecompl}

\usepackage{graphicx}	
\usepackage{amsmath}	
\usepackage{amssymb}	
\usepackage{multicol}        
\usepackage{bm}		
\usepackage{color}

\title[Pulsating A-F stars]
{Gaia luminosities of pulsating A-F stars in the Kepler field}
\author[L.A. Balona]
{L. A. Balona\\
South African Astronomical Observatory, P.O. Box 9, Observatory, Cape
2735, South Africa
}

\begin{document}

\date{Accepted .... Received ...}

\pagerange{\pageref{firstpage}--\pageref{lastpage}} \pubyear{2011}

\maketitle

\label{firstpage}

\begin{abstract}
All stars in the {\it Kepler} field brighter than 12.5 magnitude have been
classified according to variability type. A catalogue of $\delta$~Scuti and 
$\gamma$~Doradus stars is presented.  The problem of low frequencies in
$\delta$~Sct stars, which occurs in over 98\,percent of these stars, is
discussed. Gaia DR2 parallaxes were used to obtain precise luminosities, 
enabling the instability strips of the two classes of variable to be 
precisely defined.  Surprisingly, it turns out that the instability region of 
the $\gamma$~Dor stars is entirely within the $\delta$~Sct instability
strip.  Thus $\gamma$Dor stars should not be considered a separate class 
of variable.  The observed red and blue edges of the instability strip 
do not agree with recent model calculations.   Stellar pulsation occurs in
less than half of the stars in the instability region and arguments are
presented to show that this cannot be explained by assuming pulsation at a
level too low to be detected.  Precise Gaia DR2 luminosities of 
high-amplitude $\delta$~Sct stars (HADS) show that most of these are
normal $\delta$~Sct stars and not transition objects.  It is argued that
current ideas on A star envelopes need to be revised.
\end{abstract}


\section{Introduction}

The $\delta$~Sct stars are A and early F dwarfs and giants with multiple 
frequencies in the range 5--50\,d$^{-1}$ while $\gamma$~Dor stars are F 
dwarfs and giants pulsating in multiple frequencies in the range 
0.3-- 3\,d$^{-1}$.  The two types of variable have been considered as two 
separate classes of pulsating star driven by different mechanisms: 
the opacity-driven $\kappa$~mechanism in $\delta$~Sct stars and the
convective blocking mechanism in $\gamma$~Dor stars as described by 
\citet{Guzik2000}. 

Simultaneous low-frequency $\gamma$~Dor and high-frequency $\delta$~Sct 
pulsations in the same star were first discovered by \citet{Handler2002b} in
the A9/F0V star HD\,209295. Until then none of the hundreds of known
$\delta$~Sct stars were found to contain low frequencies.  The discovery was 
at variance  with the predictions of models using the $\kappa$~mechanism in
which frequencies below 5\,d$^{-1}$ are stable.  

Before the advent of the {\it CoRoT} and {\it Kepler} missions, only six 
$\delta$~Sct/$\gamma$~Dor hybrids had been discovered.  They all lie roughly 
at the high-temperature end of the known $\gamma$~Dor instability strip, but 
extending beyond the strip to higher temperatures.  With the first release of 
the {\it Kepler} data, it became clear that the hybrids were not rare 
at all \citep{Grigahcene2010}.  From only 50 days of data, at least 
one-quarter of $\delta$~Sct stars were found to be hybrids.  In this paper
it is found that significant low-frequency peaks are present in at least
98\,percent of stars.  Hybrid behaviour is the norm and occurs even among the 
hottest stars, as shown by \citet{Balona2015d}.  

The reason why so few hybrids were discovered from the ground must be partly
attributed to the fact that ground-based photometry is greatly affected by 
variations in atmospheric extinction and daily data gaps, masking the
low amplitudes of the long-period pulsations.  The few hybrids discovered
from the ground appear to be located in the region of instability where
largest amplitudes tend to occur \citep{Balona2014a}.

Recent pulsation models using time-dependent perturbation theory show that 
there is a complex interplay of driving and damping processes which cannot be 
reduced to just the $\kappa$ or convective blocking mechanisms.
\citet{Xiong2015} and  \citet{Xiong2016} show that factors such as turbulent 
dissipation, turbulent diffusion and anisotropy of turbulent convection need 
to be considered.  Furthermore, in a region with a radiative flux gradient, 
the flux will itself be modulated by the oscillations.  This is called 
``radiative modulation excitation'' (RME) by \citet{Xiong1998b}.  These
additional factors co-exist and appear to explain the low frequencies seen in
these stars.  As \citet{Xiong2016} point out, from this point of view both
$\delta$~Sct and $\gamma$~Dor stars may be regarded as a single class of
pulsating variable.

High-precision {\it Kepler} photometry has enabled a large number of pulsating
variables to be detected in the {\it Kepler} field. In this paper a list of 
$\delta$~Sct and $\gamma$~Dor stars, complete to {\it Kepler} magnitude 
$K_p = 12.5$\,mag but including fainter stars, is presented.  Using
photometric or spectroscopic estimates of effective temperature and
luminosities determined from the parallaxes in the second data release of 
Gaia (Gaia DR2; \citealt{Gaia2018}), these stars are precisely located in the 
Hertzsprung-Russell (H-R) diagram.  Comparison between the observed and
calculated red and blue edges are made.  In addition to the problem of low
frequencies in $\delta$~Sct stars, it is shown that less than half the stars
in the instability strip pulsate.  It is argued that this cannot be a result
of pulsations at a level too low to be detected.  Finally, it is shown that
most of the high-amplitude $\delta$~Sct stars (HADS) are normal $\delta$~Sct
stars and not objects intermediate between $\delta$~Sct and Cepheid
variables.

\section{The problem of the low frequencies}

Models using the $\kappa$~mechanism show pulsational instability  in
main sequence and giant stars of intermediate mass only for frequencies
higher than about 5\,d$^{-1}$.  Lower frequencies are all stable.  It may be
possible to account for low frequencies as a result of rotational splitting.
One can introduce rotational splitting in a simple way by using the know
distribution of equatorial velocities among A/F dwarfs and giants using
frequencies obtained from non-rotating models.  The frequency distributions
obtained in this way can be compared to observations, but they do not agree 
with the observed distributions \citep{Balona2015d}.  It seems that low
frequencies cannot be explained as a result of rotation.

It is also possible that inertial modes, in particular r modes
\citep{Papaloizou1978}, might account for the low frequencies.   These modes
consist of predominantly toroidal motions which do not cause compression or
expansion and hence no light variations.  However, in a rotating star, the 
toroidal motion couples with spheroidal motion caused by the Coriolis force,
leading to temperature perturbations and hence light variations.  These
modes have been recently proposed as an explanation of the broad hump that
appears just below the rotation frequency in many A stars and also period
spacings in some $\gamma$~Dor stars \citep{Saio2018a}. 

If r modes are responsible for the low frequencies in $\delta$~Sct stars, 
then all rotating stars  within the instability strip should show such 
frequencies.  This is not the case; in fact the majority of stars in the
$\delta$~Sct instability strip do not seem to pulsate at all.  For this
reason, inertial modes can be ruled out as the cause of the low frequencies
in $\delta$~Sct stars.

\citet{Balona2015d} examined the possibility that the opacities in the outer
layers of A stars may be underestimated.  Artificially increasing the
opacities by a factor of two does lead to instability of some low-degree modes
at low frequencies, but also decreases the frequency range of $\delta$~Sct
pulsations to some extent.  An increase in opacities by such a large factor
is unlikely and at present cannot be regarded as a possible solution to this
problem.

A fundamental obstacle to our understanding of stellar pulsations is that we
lack a suitable theory of convection.  The treatment of convection in 
pulsating stars has progressed quite considerably since the description of the 
$\gamma$~Dor pulsation mechanism in terms of ``convective blocking'' 
\citep{Guzik2000}.  Convective blocking uses the simplest description of 
convection and does not take into account the interaction between pulsation
and convection. Such ``frozen-in'' convection precludes the possibility of 
predicting the red edge of the  $\delta$~Sct and $\gamma$~Dor instability 
strips.  More recent treatments of pulsation use time-dependent perturbation 
theory \citep{Dupret2005,Dupret2005b}.  This allows the interaction between
pulsation and other processes, such as turbulent pressure and turbulent 
kinetic energy dissipation, to be included.

According to the time-dependent convection model of \citet{Houdek2000}, the 
damping of pulsations at the red edge of the  $\delta$~Sct instability strip 
appears to be mostly due to fluctuations of the turbulent pressure which 
oscillates out of phase with the density fluctuations.  However, in the models 
of \citet{Xiong1989} and \citet{Dupret2005}, turbulent pressure driving and 
turbulent kinetic energy dissipation damping cancel near the red edge and 
stability is determined by the perturbations of the convective heat flux.  
All models are able to predict the red edge, but further research is necessary 
to identify the correct processes.  A more detailed discussion can be found in 
\cite{Houdek2015}.

Existing theories of convection rely on unknown parameters to characterize 
the effects of turbulent pressure and turbulent kinetic energy dissipation.  
The parameters are adjusted to obtain best agreement with observations.  For 
example, three parameters are introduced by \citet{Xiong2015}.  In order to 
fix these parameters the observed red and blue edges of the $\delta$~Sct and 
$\gamma$~Dor instability strips need to be determined.  At present these are 
poorly know due to the large error in the luminosities.  This problem can now 
be solved using distances derived from Gaia DR2 parallaxes \citep{Gaia2016}.

\section{The data}

The {\it Kepler} observations consist of almost continuous photometry
of many thousands of stars over a four-year period.  The vast majority of 
stars were observed in long-cadence (LC) mode with exposure times of about 
30\,min. {\it Kepler} light curves are available as uncorrected simple 
aperture photometry (SAP) and with pre-search data conditioning (PDC) in 
which instrumental effects are removed \citep{Stumpe2012, Smith2012}. 
Most stars in the {\it Kepler} field have been observed by multicolour
photometry, from which effective temperatures, surface gravities, metal 
abundances and stellar radii can be estimated.  These stellar parameters are 
listed in the {\it Kepler Input Catalogue} (KIC, \citealt{Brown2011a}).  

Subsequently, \citet{Pinsonneault2012a} and \citet{Huber2014} revised these 
parameters for stars with $T_{\rm eff} < 6500$\,K.  For hotter stars, the KIC 
effective temperatures were compared with those determined from 
high-dispersion spectroscopy by \citet{Balona2015d}. It was found that the 
KIC temperatures are very well correlated with the spectroscopic temperatures,
but 144\,K cooler.  Adding 144\,K to the KIC temperatures reproduces the 
spectroscopic temperatures with a standard deviation of about 250\,K.  This 
can be taken as a realistic estimate of the true standard deviation since the 
KIC and spectroscopic effective temperatures are independently determined.  
In this paper the values of $T_{\rm eff}$ given by \citet{Huber2014} are used 
for stars with $T_{\rm eff} < 6500$\,K.  For hotter stars, the KIC effective 
temperatures, increased by 144\,K, are used.

To determine the luminosities of {\it Kepler} $\delta$~Sct and $\gamma$~Dor
stars requires knowledge of the apparent magnitude, interstellar extinction,
bolometric correction and the parallax.  A table of the bolometric
correction, BC, in the Sloan photometric system as a function of $T_{\rm
eff}$ and $\log g$ is presented in \citet{Castelli2003}.  For this purpose,
the small corrections described by \citep{Pinsonneault2012a} are applied to 
the {\it Kepler} $griz$ magnitudes to bring them into agreement with the 
Sloan system.   Correction for interstellar extinction was applied to the
$r$ magnitude using $r_0 = r - 0.874A_V$ \citep{Pinsonneault2012a}.

The value of $A_V$ listed in the KIC is from a simple reddening model which
depends only on galactic latitude and distance. A three-dimensional
reddening map with a radius of 1200\,pc around the Sun and within 600\,pc of
the galactic midplane has been calculated by \citet{Gontcharov2017}.  This is
likely to produce more accurate values of $A_V$ and is used in this paper. 
For more distant stars, the simple reddening model is used but adjusted so
that it agrees with the 3D map at 1200\,pc.  A comparison shows that the
KIC values of $A_V$ are typically 0.017 mag higher than those given by the 
3D map.

From the Gaia DR2 parallax, $\pi$, the absolute magnitude is calculated using 
$r_{\rm abs} = r_0 + 5(\log_{10}\pi + 1)$.  The absolute bolometric magnitude 
is then given by $M_{\rm bol} = r_{\rm abs} + {\rm BC}_r - M_{\rm bol\odot}$ 
with the solar absolute bolometric magnitude $M_{\rm bol\odot} = 4.74$.  
Finally, the luminosity relative to the Sun is found using $\log L/L_\odot = 
-0.4M_{\rm bol}$.

\section{Classification and light curves}

Stars in the {\it Kepler} field were observed almost continuously for 17
quarters covering a period of just over 4 years.  Light curves and
periodograms of all short-cadence observations (4827 stars) were visually
examined.  All long-cadence data brighter than magnitude 12.5, but including
many more stars fainter than this limit (20784 stars in total) were visually 
examined as well.

Detection of $\delta$~Sct stars is relatively easy since the periodograms
show peaks at high frequencies.  The $\beta$~Cep variables and some types
of compact stars also show high frequencies, but these can be distinguished 
from $\delta$~Sct using the KIC effective temperatures and surface gravities.
Among the 20784 stars examined, 1740 $\delta$~Sct stars were discovered.

The distinction between $\delta$~Sct and $\gamma$~Dor stars was based purely
on the absence of peaks with significant amplitudes having frequencies in
excess of around 5\,d$^{-1}$.  These stars have multiple peaks below this
frequency.  

It is sometimes difficult to distinguish between $\gamma$~Dor
and rotating variables which are the very common.  Classification as 
a $\gamma$~Dor star was made only if the frequencies were too widely spread 
to be due to differential rotation.  It should be noted that surface 
differential rotation reaches a maximum in the F stars \citep{Balona2016b}
where most $\gamma$~Dor stars are to be found.  It is possible that at least 
some frequency peaks in $\gamma$~Dor stars may be due to rotation.
$\gamma$~Dor stars are distinguished from the slowly pulsating B (SPB) stars 
and some compact objects using the KIC effective temperatures and surface 
gravities.  Among the 20784 stars examined, 820 $\gamma$~Dor stars were 
discovered.

During the course of examination of the periodograms, instances were noted
of the presence of low-frequency peaks in $\delta$~Sct stars, excluding
peaks which might be attributed to binarity or rotation.  The number of
stars without significant low frequencies was found to be very low, probably
less than 2\,percent of the $\delta$~Sct stars.  Thus nearly all
$\delta$~Sct stars are hybrids.

Even a cursory examination of the {\it Kepler} light curves reveals a set of 
stars with beating and highly asymmetric minima and maxima.  Maximum light 
amplitudes far exceeded those of minimum light.  In many cases sudden 
high-amplitude excursions can easily be mistaken for flares.  The morphology of 
these light curves is striking and quite unlike any other type of variable.  
These $\gamma$~Dor stars were first described by \citet{Balona2011f} who 
named them the ASYM (asymmetric) type of $\gamma$~Dor variable.

\begin{figure}
\centering
\includegraphics[]{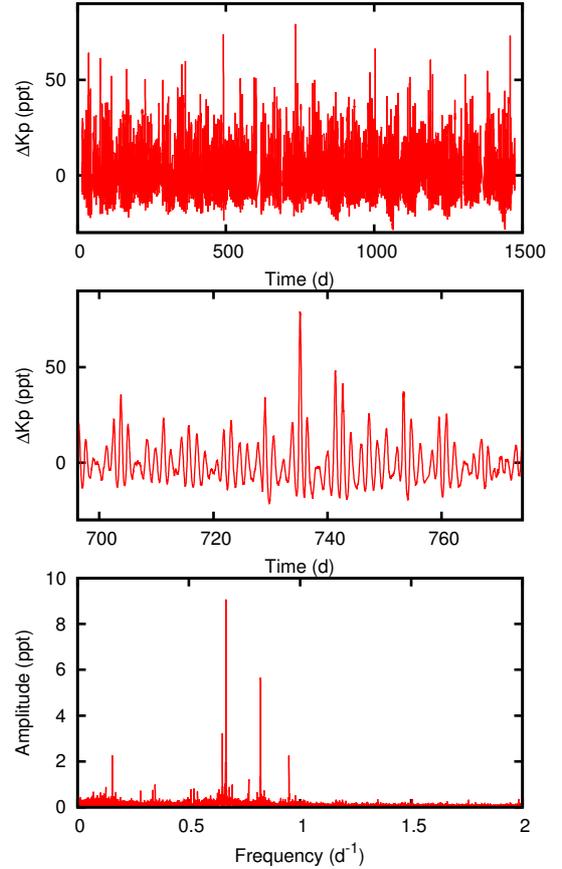}
\caption{Top panels: {\it Kepler} light curve of KIC\,8180361 showing
asymmetrical light curve with occasional large excursions, one of which is
depicted in the middle panel.  The periodogram is shown in the bottom panel.}
\label{gdora}
\end{figure}

Fig,\,\ref{gdora} shows an example of the light curve and periodogram of the
ASYM type.  It should be noted that the {\it Kepler} PDC light curve flags
most of the flare-like excursions as bad points.  The light curve shown in
the figure was reconstructed from the raw data using the good points of the
PDC data to determine the necessary corrections.  One of the questions that 
need to be asked is whether the large excursions arise as a result of beating 
of sinusoidal components.  This can be answered by clipping the light curve 
to eliminate the large excursions.  If the sudden high maxima are a simple
result of beating of pure sinusoids, clipping these maxima should not
introduce new frequencies.  The periodogram of the clipped data shows 
significantly fewer low-amplitude peaks, indicating that these additional 
frequency components are required and that the excursions are simply due to a 
highly non-linear physical process.  

The light curves of most $\gamma$~Dor stars show characteristic beating, but 
with symmetric minima and maxima (the SYM type). The beating may be traced to 
two or more dominant closely-spaced frequencies in the periodograms.  Another 
type of $\gamma$~Dor star shows no obvious beating in the light curve and an 
even frequency spread of peaks in the periodogram with comparable amplitudes 
(the MULT type).  Because the distinction between the three groups may be 
important, each star was classified as either GDORA,GDORS or GDORM 
corresponding to the ASYM, SYM and MULT types.  

Among the 820 $\gamma$~Dor stars there are 137 GDORA and 447 GDORS stars.
There are 16 stars which are GDORS for some of the time and GDORA at  other 
times.  There are 215 GDORM stars.  A few stars which are difficult to 
classify into the three groups are labeled simply as GDOR.

\begin{table}
\begin{center}
\caption{An extract of the $\delta$~Sct and $\gamma$~Dor catalogue in the 
{\it Kepler} field.  The full table is available in electronic form. The
first two columns is the KIC number and the type of variable.  The Gaia DR2 
parallax (Plx) and its standard deviation (e\_Plx) are in milliarcseconds.  
The effective temperature and its error are in K.  The interstellar
absorption, $A_V$ (mag) is derived from the tables in \citet{Gontcharov2017}. 
Finally, the relative solar luminosity and its error is given.}
\label{table1}
\resizebox{8.5cm}{!}{
\begin{tabular}{llrrrrrrr}
\hline
\multicolumn{1}{c}{KIC}                  & 
\multicolumn{1}{c}{Type}                 &
\multicolumn{1}{c}{Plx}                  & 
\multicolumn{1}{c}{e\_Plx}               &
\multicolumn{1}{c}{$T_{\rm eff}$}        & 
\multicolumn{1}{c}{e\_$T_{\rm eff}$}     &
\multicolumn{1}{c}{$A_V$}                &
\multicolumn{1}{c}{$\log{L/L_\odot}$}    & 
\multicolumn{1}{c}{e\_$\log{L/L_\odot}$} \\
\hline
  1026294 & DSCT           &  1.0562 &  0.0245 &  8083 &   280 & 0.38 & 1.2026 &  0.042 \\
  1161908 & GDORS          &  0.8361 &  0.0141 &  6662 &   198 & 0.45 & 0.8021 &  0.041 \\
  1162150 & DSCT           &  0.9272 &  0.0247 &  7015 &   263 & 0.46 & 1.6332 &  0.042 \\
  1163943 & DSCT           &  1.2370 &  0.0317 &  7236 &   241 & 0.33 & 1.2845 &  0.042 \\
  1294670 & DSCT           &  0.5840 &  0.0254 &  7220 &   276 & 0.53 & 1.6572 &  0.044 \\
  1430590 & DSCT           &  0.8833 &  0.0278 &  6899 &   253 & 0.48 & 1.0904 &  0.042 \\
  1430741 & GDORM          &  0.4982 &  0.0171 &  7091 &   277 & 0.53 & 1.1261 &  0.043 \\
  1431379 & GDORS          &  1.0646 &  0.0242 &  7106 &   245 & 0.39 & 0.9482 &  0.042 \\
  1431794 & DSCT/ROT       &  0.8712 &  0.0234 &  7255 &   273 & 0.46 & 1.2349 &  0.042 \\
\hline
\end{tabular}
}
\end{center}
\end{table}

\begin{table}
\begin{center}
\caption{An extract of the catalogue of $\delta$~Sct and $\gamma$~Dor
stars in the {\it Kepler} field for which Gaia DR2 luminosities could not be
obtained.  The {\it Kepler} magnitude, $K_p$, the effective temperature,
$T_{\rm eff}$, and its error and the luminosity, $\log L/L_\odot$, determined 
from $T_{\rm eff}$ and the KIC radii are given.  The error in
$\log(L/L_\odot)$ is about 0.4 dex.}
\label{table2}
\begin{tabular}{llrrrr}
\hline
\multicolumn{1}{c}{KIC}                  & 
\multicolumn{1}{c}{Type}                 &
\multicolumn{1}{c}{Kp}                  & 
\multicolumn{1}{c}{$T_{\rm eff}$}        & 
\multicolumn{1}{c}{e\_$T_{\rm eff}$}     &
\multicolumn{1}{c}{$\log{L/L_\odot}$}    \\
\hline
  1571152 & DSCT    &            9.268 &  7192 &  149 &  1.88 \\
  2568519 & GDORS   &           11.258 &  6299 &  182 &  0.18 \\
  2572386 & DSCT    &           13.278 &  7345 &  275 &       \\
  2856756 & DSCT    &           10.250 & 10477 &  365 &  2.39 \\
  2975832 & DSCT    &           12.610 &  6824 &  249 &       \\
\hline
\end{tabular}
\end{center}
\end{table}

A catalogue of the $\delta$~Sct and $\gamma$~Dor stars is available in
electronic form.  An extract from the catalogue is shown in 
Table\,\ref{table1}.  In this table the effective temperatures for 
$T_{\rm eff} < 6500$\,K are from \citet{Huber2014}.  For hotter stars, 144\,K 
has been added to the KIC effective temperature as discussed above. The 
interstellar absorption, $A_V$, is obtained from the 3D map of 
\citet{Gontcharov2017}. If the KIC values of $A_V$ are used, the higher 
absorption leads to a slight increase in $\log L/L_\odot$ of only 
0.006 dex.  There are 1680 $\delta$~Sct stars and 796 $\gamma$~Dor stars with
luminosities estimated from Gaia DR2 parallaxes.  

For some stars Gaia DR2 parallaxes do not exist, no effective temperature
is available or the interstellar extinction cannot be estimated.  These 
94 stars are listed separately (see Table\,\ref{table2} for an extract).  

The mean difference between the photometrically estimated luminosities, 
$\log(L/L_\odot)_P$, and the luminosities from Gaia DR2, $\log(L/L_\odot)_G$, 
is $<\log(L/L_\odot)_P -\log(L/L_\odot)_G> = -0.22 \pm 0.01$.  The
photometrically estimated luminosities have a standard error of 0.39 dex.  

When deriving the location of the instability strips, the surface
gravity, $\log g$, as a function of $T_{\rm eff}$ is sometimes used instead
instead of $\log L/L_\odot$ as a function of $T_{\rm eff}$ (eg.
\citealt{Uytterhoeven2011}). This is simply because $\log g$ can be directly
obtained from the observations.  However, for comparison with theoretical
models, $\log L/L_\odot$ is to be preferred.  In this paper, $\log L/L_\odot$ 
is the natural choice because it can be obtained directly from the parallax 
as described above.  Moreover, the Gaia DR2 parallaxes result in luminosities 
with very high accuracy, so that the instability strip can be determined 
far more precisely than the use of $\log g$.

Many stars are binaries.  If a star is a binary with components of equal 
luminosity, the luminosity calculated from the parallax will be twice as 
large as a single star of the same luminosity.  Thus $\log L/L_\odot$ will be 
too high by about 0.3 dex.  We do not know which stars are binaries in the 
{\it Kepler} field and it is possible that the estimated $\log L/L_\odot$ may 
be too large for some stars.  If the components have different temperatures, 
this will also affect the $T_{\rm eff}$ of the combined stars.  Unfortunately 
without detailed spectroscopic observations of each star it is impossible to 
correct for these effects.

\section{The $\delta$~Sct stars}

\begin{figure}
\centering
\includegraphics[]{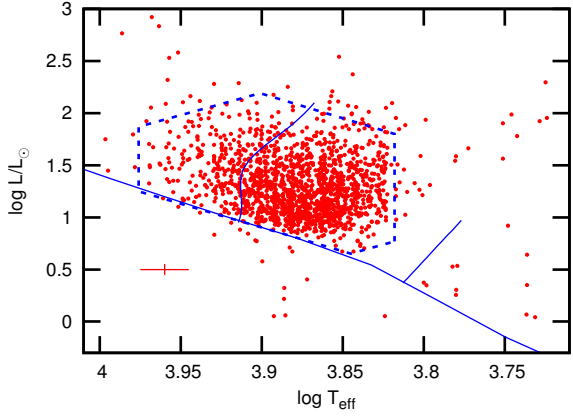}
\caption{{\it Kepler} $\delta$~Sct stars in the H-R diagram (dots) using
Gaia DR2 parallaxes and \citet{Gontcharov2017} values of $A_V$.  The solid 
lines are the zero-age main sequence (solar abundance from 
\citet{Bertelli2008}) and the nonradial red and blue edges from 
\citet{Xiong2016}.  The dashed polygon defines the region which includes the 
majority of $\delta$~Sct stars.  The cross on the bottom left shows the 
1-$\sigma$ error bars.}
\label{dsct}
\end{figure}

In Fig.\,\ref{dsct} the $\delta$~Sct stars are shown in the H-R diagram
together with the zero-age main sequence from models with solar abundances
($Z = 0.017$) and helium abundance $Y = 0.26$  by \citet{Bertelli2008}.  The 
dashed polygon is a visual estimate of the location of the majority of the 
$\delta$~Sct stars and includes 94\,percent of these stars.  Most of the stars 
are within the temperature range $6580 < T_{\rm eff} < 9460$\,K. The typical 
standard deviation is about 260\,K in $T_{\rm eff}$ (0.015 in 
$\log T_{\rm eff}$) and about 0.052 in $\log(L/L_\odot)$.  These errors are 
shown by the cross in Fig.\,\ref{dsct}.  The main uncertainty in the 
luminosities is the effect of interstellar light absorption.  The extinction 
values used here are interpolated from the table by \citet{Gontcharov2017}.
Reduced extinction will lead to smaller luminosities.   Also shown in the 
figure are the red and blue edges from \citet{Xiong2016}.   The effective 
temperatures of the red and blue edges are clearly too cool.

\begin{figure}
\centering
\includegraphics[]{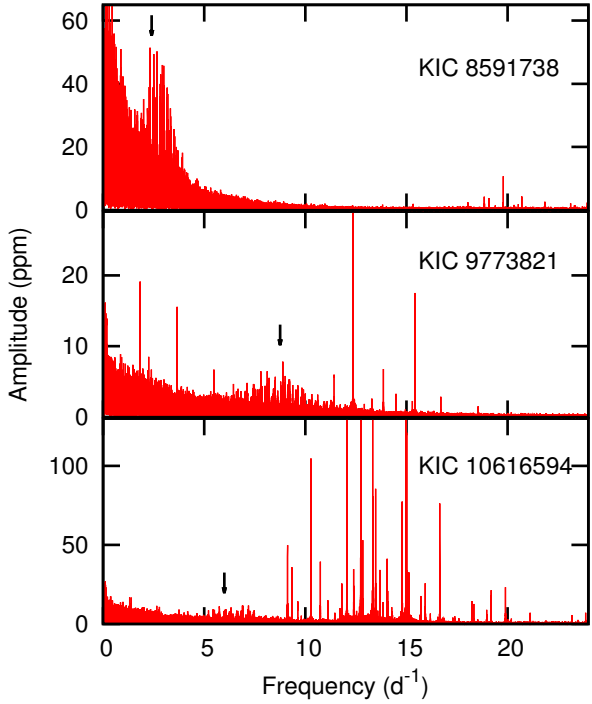}
\caption{Examples of stars which show both solar-like oscillations
(indicated by the arrow) and $\delta$~Sct high frequencies.  These are
composite stars containing a cool giant and a $\delta$~Sct star.}
\label{dsol}
\end{figure}

There are a number of outliers on both the hot and cool sides of the
instability region which need further study.  It is possible that their
effective temperatures are in error, but a study by \citet{Balona2016c}
suggests that there is evidence for a class of variables with  multiple
high frequencies characteristic of $\beta$~Cep and $\delta$~Sct stars which
lie between the red edge of the $\beta$~Cep and and the blue edge of the
$\delta$~Sct instability regions.  These have been called Maia variables.
The few outliers below the ZAMS may be evolved objects, though the KIC 
surface gravities seem to be normal.

\citet{Qian2018} observed a group of 131 cool multiperiodic variable stars 
that are much cooler than the red edge of the $\delta$~Sct instability
strip.  Many of these are in the {\it Kepler} field.  Inspection of
their periodograms show the typical Gaussian amplitude envelope
characteristic of solar-like oscillations, so it is possible that
\citet{Qian2018} have mis-classified solar-like pulsations in red giants as
$\delta$~Sct stars.  There is no indication of a cool population of 
$\delta$~Sct stars among the {\it Kepler} data examined.   There are, however,
composite objects consisting of a cool giant and a normal $\delta$~Sct star. 
Fig.\,\ref{dsol} shows periodograms of three of these stars where the 
Gaussian-like amplitude envelope characteristic of solar-like oscillations in 
a cool giant and the pulsations in a $\delta$~Sct star are clearly visible.

There are other cool stars that can be classified as $\delta$~Sct variables.
An example is KIC\,4142768 where the effective temperature from several 
sources (including the KIC) is only about 5400\,K.  The star is a heartbeat 
variable \citep{Balona2018b} and the LAMOST spectrum is A9V with no sign of a 
cool giant.  In this case the explanation may be additional reddening caused  
by gas and dust associated with the binary.

\begin{figure}
\centering
\includegraphics[]{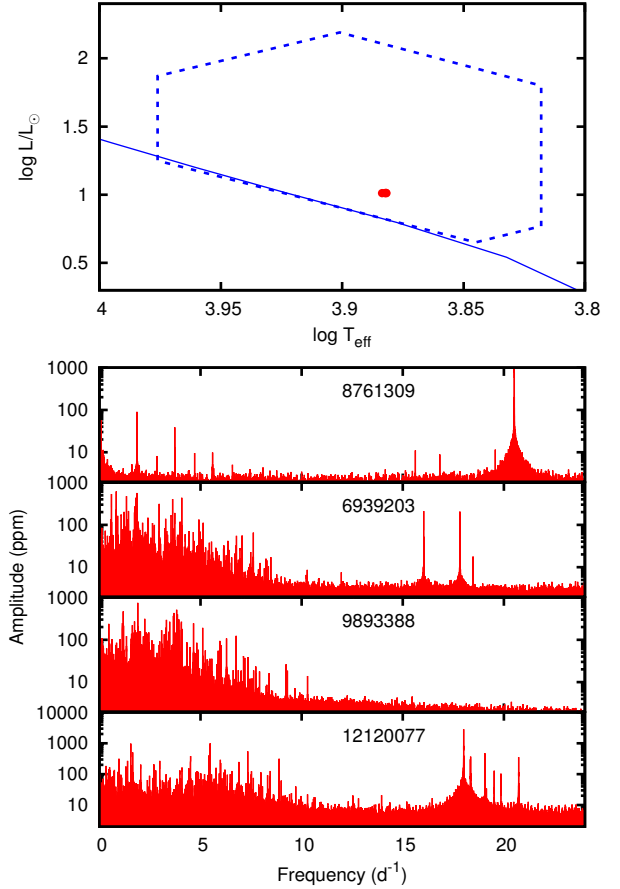}
\caption{Example of periodograms of four $\delta$~Sct stars which have 
practically the same stellar parameters.  The top panel shows their location
in the H-R diagram, all within the single filled circle.  The ZAMS and the
$\delta$~Sct instability region are shown.}
\label{simlc}
\end{figure}

\begin{figure}
\centering
\includegraphics[]{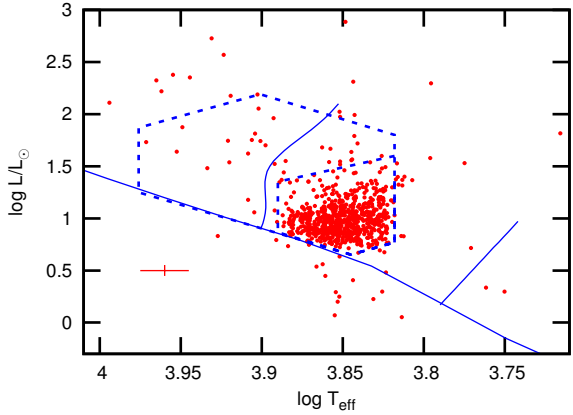}
\caption{The location of the $\gamma$~Dor stars in the H-R diagram (dots) 
using Gaia DR2 parallaxes and \citet{Gontcharov2017} values of $A_V$.  The
outer dashed polygon defines the instability region of $\delta$~Sct stars. 
The smaller polygon defines the approximate location of $\gamma$~Dor stars.  
The solid lines show the zero-age main sequence and the red and blue edges 
of the instability strip calculated by \citet{Xiong2016}.}
\label{gdor}
\end{figure}

\begin{table}
\begin{center}
\caption{Coordinates of the vertices of the $\delta$~Sct and $\gamma$~Dor
instability regions as shown in Figs.\,\ref{dsct} and \ref{gdor}.}
\label{boxes}
\begin{tabular}{rrrr}
\hline
\multicolumn{2}{c}{$\delta$~Sct} & 
\multicolumn{2}{c}{$\gamma$~Dor} \\
\multicolumn{1}{c}{$\log T_{\rm eff}$}&
\multicolumn{1}{c}{$\log L/L_\odot$} & 
\multicolumn{1}{c}{$\log T_{\rm eff}$}&
\multicolumn{1}{c}{$\log L/L_\odot$} \\
\hline
  3.818 & 0.770  &      3.818 & 1.600 \\
  3.818 & 1.800  &      3.890 & 1.350 \\
  3.901 & 2.190  &      3.890 & 0.867 \\
  3.976 & 1.870  &      3.845 & 0.651 \\
  3.976 & 1.250  &      3.818 & 0.770 \\
  3.845 & 0.651  &      3.818 & 1.400 \\
  3.818 & 0.770  &            &       \\
\hline
\end{tabular}
\end{center}
\end{table}

Fig.\,\ref{simlc} is an illustration of the disparity in frequencies and 
amplitudes among $\delta$~Sct stars with practically the same stellar 
parameters.  All four stars have $\log T_{\rm eff} = 7630$\,K and 
$\log L/L_\odot = 1.012$ within 20\,K and 0.002 dex respectively.  The
disparity in the general appearance in frequency peaks is remarkable.  It is
possible that rotation may be an important factor or that some of the four 
stars may be composite which will affect the derived luminosity.  Also, the 
observational error of around 250\,K in $T_{\rm eff}$ could modify the 
expected pulsation frequencies somewhat.  Nevertheless, the general
impression obtained from visual inspection is that the periodogram of each
star is unique.

\section{The $\gamma$~Dor stars}

In Fig.\,\ref{gdor} the $\gamma$~Dor stars are shown in the H-R diagram
together with the zero-age main sequence from models with solar abundance
\citep{Bertelli2008}.  The figure shows the instability polygon of the
$\delta$~Sct stars as reference.  The smaller nested polygon contains
89\,percent of the $\gamma$~Dor stars.  The coordinates of the vertices in
the polygonal regions of the $\delta$~Sct and $\gamma$~Dor stars are listed
in Table\,\ref{boxes}.

As mentioned above, a star was classified as a $\gamma$~Dor variable only if
all the peaks in the periodogram are below 5\,d$^{-1}$ (with some leeway if
there are a few peaks of low amplitude above this frequency).  In addition,
a very important qualification is added: the frequencies must exclude
rotational modulation.  If, for example, a peak and its harmonic is seen,
then it is a rotational variable and not a $\gamma$~Dor. It is this
criterion more than anything else which is responsible for refining the
$\gamma$~Dor instability strip.  Of course, the luminosities derived from
Gaia DR2 also assist in this refinement.  The location of $\delta$~Sct and
$\gamma$~Dor stars using KIC radii and effective temperatures to determine
the luminosities is shown in Fig.\,2 of \citet{Balona2014a}.  These can be
compared with Fig.\,\ref{dsct} and \ref{gdor} in this paper.

It is interesting that the $\gamma$~Dor instability region lies completely
within the $\delta$~Sct instability region.  In the $\gamma$~Dor box there 
are 711 $\gamma$~Dor stars, but there are also 815 $\delta$~Sct stars (and 
994 other objects) in the same box.  It seems that the $\gamma$~Dor
variables are simply a subset of the $\delta$~Sct stars and not an
independent class, as suggested by \citet{Xiong2016}.  There is no difference 
in the locations of the GDORA, GDORS and GDORM subtypes within the 
$\gamma$~Dor box.  The  red and blue edges of the $\gamma$~Dor instability 
strip calculated by \citet{Xiong2016} are shown in the figure, but do not 
agree with observations.

There are quite a number of hot outliers which have been studied by
\citet{Balona2016c}.  The spectra confirm that many of these stars are
indeed hot $\gamma$~Dor stars.  Whether or not these deserve a separate
classification or whether these stars, like the $\gamma$~Dor stars,
may just be due to unusual mode selection processes remains to be seen.

\citet{Mowlavi2013} found a large population of new variable stars between
the red edge of the SPB stars and the blue edge of the $\delta$~Sct stars, a 
region in the H-R diagram  where no pulsation is predicted to occur based on 
standard stellar  models.  Their periods range from 0.1--0.7\,d, with
amplitudes  between 1 and 4\,mmag.  It is possible that these could be
identified with the hot $\gamma$~Dor stars in the {\it Kepler} field.

The three coolest $\gamma$~Dor stars (KIC\,4840401, 8264287, and 12218727)
are not solar-like variables.  No  known variable class in this temperature
range resembles the $\gamma$~Dor class.  They could be composite objects,
but merit further study.  The stars lying below the ZAMS also merit further
study.  They may perhaps be evolved compact objects.

\section{Fraction of pulsating stars}

\begin{figure}
\centering
\includegraphics[]{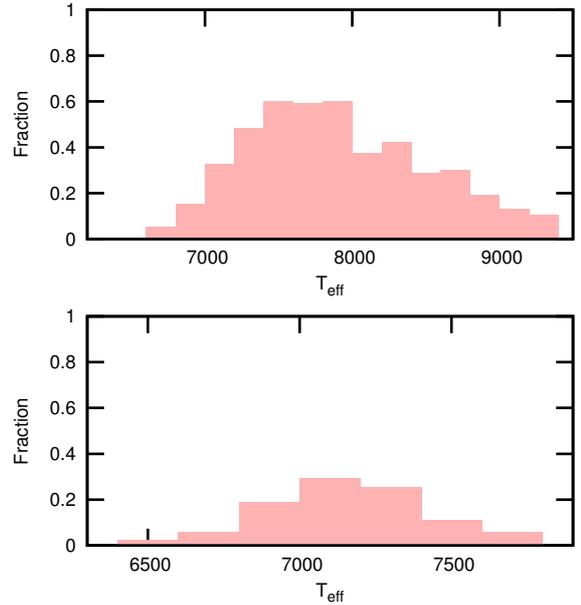}
\caption{Top panel: the fraction of $\delta$~Sct stars in the instability
box relative to all stars in the box as a function of effective temperature.  
Bottom panel: the fraction of $\gamma$~Dor stars relative to all stars in
the $\gamma$~Dor box as a function of effective temperature.}
\label{dsctdis}
\end{figure}

In order to determine whether or not all stars in the $\delta$~Sct
instability region pulsate, it is necessary to count the number of pulsating
and non-pulsating stars within the same region. To assure completeness, the 
sample is limited to stars with $K_p < 12.5$\,mag because all stars in the 
{\it Kepler} field down to this brightness level have been classified 
according to variability class.

It is found that 2881 stars for which luminosities can be determined from
Gaia DR2 lie within the $\delta$~Sct instability box (excluding known evolved 
objects). Of these, 874 are $\delta$~Sct stars and 281 are $\gamma$~Dor
stars.  The remainder are non-pulsating as far as can be ascertained.  Most
are rotational variables or eclipsing systems.  There are 1759 stars
within the $\gamma$~Dor instability region, of which 260 are $\gamma$~Dor
stars and 401 are $\delta$~Sct stars. 

The number of $\delta$~Sct stars relative to the total number of stars within 
the $\delta$~Sct instability box varies as a function of effective 
temperature (Fig.\,\ref{dsctdis}, top panel).  The largest fraction of
$\delta$~Sct stars occurs around  $T_{\rm eff} \approx 7600$\,K.  The 
bottom panel of the same figure shows the relative number of $\gamma$~Dor 
stars in the $\gamma$~Dor instability box as a function of temperature.

\begin{figure}
\centering
\includegraphics[]{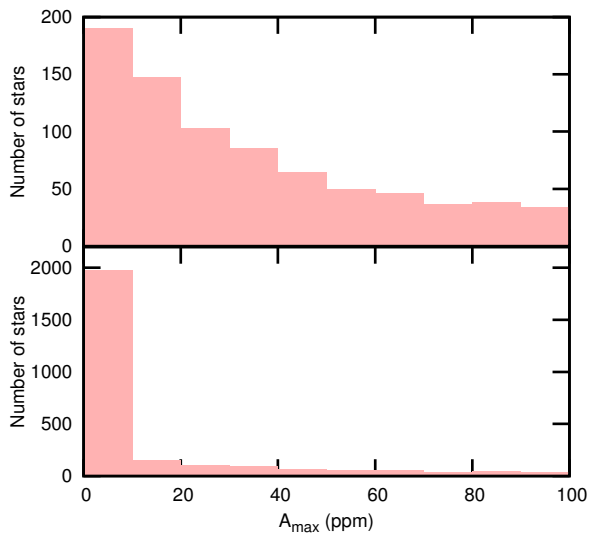}
\caption{Top panel: the distribution of the maximum amplitude (ppm) in 
{\it Kepler} $\delta$~Sct stars with $K_p < 12.5$\,mag in the instability box.  
Bottom panel: As above, but including all stars with $K_p < 12.5$\,mag in the 
$\delta$~Sct instability box on the assumption that these ``constant'' stars 
actually pulsate at below the detection limit. The peak at $A_{\rm max} < 
10$\,ppm reaches 1781 stars.}
\label{ampdist}
\end{figure}

These results strongly indicate that both pulsating and non-pulsating stars
co-exist within the instability strip.  This poses a problem because it is
difficult to understand why some stars with closely similar parameters, and 
with presumably very similar driving and damping regions, should pulsate
while others do not pulsate.    It is possible to argue that the non-pulsating 
stars do pulsate, but at a level below the detection limit, but statistics of 
the  amplitude distribution argue against this \citep{Balona2011g}.

In the top panel of Fig\,\ref{ampdist}, the distribution of maximum amplitude 
for $\delta$~Sct stars with $K_p < 12.5$\,mag is shown.  As might be expected, the 
number of stars increases as the amplitude decreases.  If we assume that the 
non-pulsating stars are actually pulsating below the detection level, they 
should also be included in this distribution.  They must then be added to the 
number in the bin with the lowest amplitude covering amplitudes between 0 and 
10\,ppm.  If that is done, there is a discontinuous jump in the amplitude
distribution, as seen in the bottom panel of Fig.\,\ref{ampdist}, which does 
not appear to be physical. 

The only way of avoiding this strange behaviour in the amplitude distribution
is to assume that the non-pulsating stars belong to a different population
and should not be included in the amplitude distribution of the pulsating
stars.  In other words, the simplest explanation is that the non-pulsating
stars are not pulsating below the detection limit and do not pulsate
at all.  It can be concluded that the $\delta$~Sct instability strip is not 
pure.  As already mentioned, this poses a serious problem.  We seem to have 
an incomplete understanding of the outer layers of A stars.

The conclusion here differs from that of \citet{Murphy2015}.  From a study of 
only 54 stars, they found that all stars within the $\delta$~Sct instability 
strip pulsate.  \citet{Guzik2014} found that most stars pulsate, but a few 
constant stars remain.  It could be argued that most of the constant stars are
outside the instability strip due to errors in the effective temperature. The 
typical error in $T_{\rm eff}$ is about 200--300 K, while the width of the 
instability strip is about 3000\,K.   To move a star in the middle of the 
instability strip to the edges of the strip requires that$T_{\rm eff}$ be in 
error by about 5 standard deviations (a probability less than $10^{-5}$).  Of 
course, the probability will be higher if the star is closer to the edge of 
the instability strip, but it means that the probability that all 1781 
non-pulsating stars are outside the instability strip is the product of the 
individual probabilities which is essentially zero.  Furthermore, one has to 
assume that (for some unknown reason) the values of $T_{\rm eff}$ for 
$\delta$~Sct stars are much more accurate,  otherwise many of these 
$\delta$~Sct stars would also be moved out of the instability region.  

For these reasons it is a certainty that non-pulsating stars exist in the
instability region unless the discontinuity in the amplitude distribution
can be understood in some other way.
 
\section{High-amplitude $\delta$~Sct stars}

The high-amplitude $\delta$~Sct stars (HADS) are a well-known group
characterized by high photometric amplitude (generally higher than 0.3\,mag)
and fairly simple frequency spectra, but with many combination frequencies. 
None of the stars in the {\it Kepler} field attain such a large amplitude,
but several are known in the general field.  They have been assumed to be
transition objects between Cepheids and $\delta$~Sct stars - in fact they
were originally called ``dwarf Cepheids''.   

It is interesting to locate the field HADS in the H-R diagram to determine
their evolutionary status.  Using the catalogue of \citet{Rodriguez2000},
all stars with amplitudes exceeding 0.3 mag were selected.  Gaia DR2 
parallaxes were obtained for those stars which have effective temperatures 
from Apsis-Priam \citep{Bailer-Jones2013}.  These effective temperatures are 
available in the Gaia DR2 catalogue. The interstellar absorption was
estimated using the 3D map of \citet{Gontcharov2017}.  Results are shown in 
Table\,\ref{tabhads}.  Some of the HADS appear to be evolved stars belonging
to Population II on the basis of their high proper motions and low
metallicities.  These are called SX Phe variables.

\begin{table}
\begin{center}
\caption{List of HADS stars with know effective temperatures.  Column two is
the type of star.  The light amplitude is from the catalogue of 
\citet{Rodriguez2000}.  The parallax is from Gaia DR2 and the interstellar 
absorption, $A_V$ mag, is from the table in \citet{Gontcharov2017}.  The 
effective temperature is from the Apsis-Priam compilation in the
Gaia DR2 catalogue and has a typical error of about 250\,K.  The last column 
is the luminosity with typical error of 0.04\,dex.}
\label{tabhads}
\resizebox{8.5cm}{!}{
\begin{tabular}{llrrrrr}
\hline
\multicolumn{1}{c}{Star}            & 
\multicolumn{1}{c}{Type}            &
\multicolumn{1}{c}{Amp}             &
\multicolumn{1}{c}{Plx}             &
\multicolumn{1}{c}{$A_V$}             &
\multicolumn{1}{c}{$T_{\rm eff}$}   &
\multicolumn{1}{c}{$\log \tfrac{L}{L_\odot}$} \\
\multicolumn{1}{c}{}            & 
\multicolumn{1}{c}{}            &
\multicolumn{1}{c}{mag}             &
\multicolumn{1}{c}{mas}             &
\multicolumn{1}{c}{mag}             &
\multicolumn{1}{c}{K}   &
\multicolumn{1}{c}{} \\
\hline 
XX Cyg      & SXPHE  & 0.80 &  $0.84 \pm 0.04$ &  0.55 & 6982 & 1.50 \\
KZ Hya      & SXPHE  & 0.80 &  $2.98 \pm 0.10$ &  0.27 & 7239 & 1.07 \\
CY Aqr      & SXPHE  & 0.71 &  $2.34 \pm 0.05$ &  0.39 & 7271 & 0.93 \\
AI Vel      & DSCT   & 0.67 &  $9.86 \pm 0.03$ &  0.31 & 6944 & 1.43 \\
RS Gru      & DSCT   & 0.56 &  $4.03 \pm 0.05$ &  0.24 & 7226 & 1.49 \\
DY Peg      & SXPHE  & 0.54 &  $2.45 \pm 0.07$ &  0.39 & 7646 & 1.14 \\
GP And      & DSCT   & 0.52 &  $1.94 \pm 0.15$ &  0.37 & 7718 & 1.14 \\
DY Her      & DSCT   & 0.51 &  $1.39 \pm 0.04$ &  0.36 & 6920 & 1.58 \\
SZ Lyn      & DSCT   & 0.51 &  $2.49 \pm 0.07$ &  0.22 & 7799 & 1.41 \\
EH Lib      & DSCT   & 0.50 &  $2.72 \pm 0.05$ &  0.37 & 7011 & 1.23 \\
VZ Cnc      & DSCT   & 0.50 &  $4.43 \pm 0.05$ &  0.22 & 6812 & 1.65 \\
V4425 Sgr   & SXPHE  & 0.49 &  $2.22 \pm 0.05$ &  0.54 & 7029 & 1.59 \\
BS Aqr      & DSCT   & 0.44 &  $1.98 \pm 0.06$ &  0.27 & 7009 & 1.69 \\
AE UMa      & SXPHE  & 0.44 &  $1.28 \pm 0.07$ &  0.24 & 7825 & 1.25 \\
CW Ser      & DSCT   & 0.43 &  $0.53 \pm 0.04$ &  0.61 & 7271 & 1.96 \\
YZ Boo      & DSCT   & 0.42 &  $1.68 \pm 0.03$ &  0.29 & 7365 & 1.36 \\
VX Hya      & DSCT   & 0.40 &  $0.99 \pm 0.04$ &  0.38 & 6694 & 1.82 \\
BE Lyn      & DSCT   & 0.39 &  $3.86 \pm 0.05$ &  0.07 & 7806 & 1.23 \\
SS Psc      & DSCT   & 0.39 &  $0.82 \pm 0.12$ &  0.41 & 7286 & 1.88 \\
ZZ Mic      & DSCT   & 0.35 &  $2.96 \pm 0.05$ &  0.26 & 7757 & 1.28 \\
RY Lep      & DSCT   & 0.35 &  $2.46 \pm 0.05$ &  0.21 & 7122 & 1.90 \\
V0567 Oph   & DSCT   & 0.33 &  $1.45 \pm 0.04$ &  1.02 & 5727 & 1.59 \\
DE Lac      & DSCT   & 0.32 &  $1.28 \pm 0.04$ &  0.59 & 6109 & 1.88 \\
V1719 Cyg   & DSCT   & 0.31 &  $2.52 \pm 0.03$ &  0.39 & 6531 & 2.09 \\
AD CMi      & DSCT   & 0.30 &  $2.18 \pm 0.04$ &  0.24 & 7129 & 1.60 \\
\hline
\end{tabular}
}
\end{center}
\end{table}

\begin{figure}
\centering
\includegraphics[]{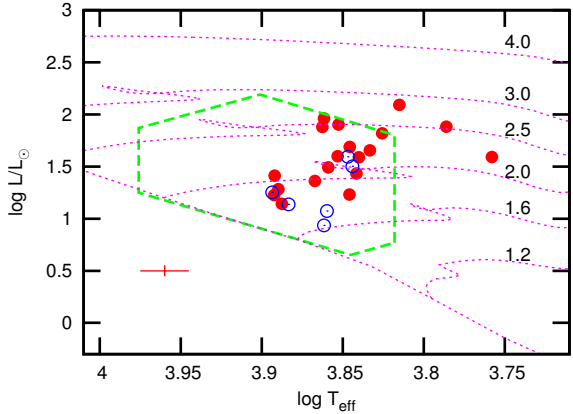}
\caption{Location of the HADS in the H-R diagram.  The filled circles are
HADS of the $\delta$~Sct type and the open circles are SX Phe stars.  The
$\delta$~Sct box, the ZAMS and evolutionary tracks (masses labeled) from
\citet{Bertelli2008} are shown.}
\label{hads}
\end{figure}

Fig.\,\ref{hads} shows the HADS in the H-R diagram.  They appear to lie in
the middle of the instability strip except for V0567 Oph which has a very
low effective temperature.  The SX Phe group also lie well within the
instability box.  If HADS were transition objects between $\delta$~Sct and
Cepheids, one would expect all of them to have high luminosities,
intermediate between the two groups of variables.  However, most HADS appear 
to be normal $\delta$~Sct stars.

\begin{figure}
\centering
\includegraphics[]{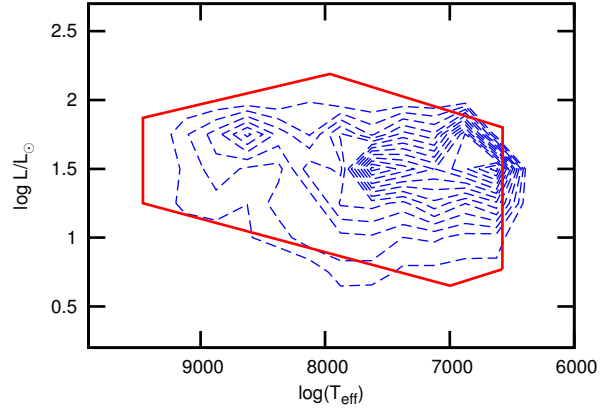}
\caption{Contours showing maximum pulsation amplitudes of $\delta$~Sct stars
in the H-R diagram with polygonal region of instability.  The contours show
amplitudes from 500 to 7000 ppm.}
\label{amp}
\end{figure}

Fig.\,\ref{amp} shows how the typical maximum amplitude in $\delta$~Sct
stars varies within the instability strip.  Largest amplitudes tend to occur
around $T_{\rm eff} \approx 7000$\,K among the more luminous cool $\delta$~Sct
stars.  The 
mean effective temperature of the HADS is $7200 \pm 200$ which suggests that 
the high amplitudes in HADS are in line with what might be expected in normal 
$\delta$~Sct stars.   The reason why the amplitude is so much larger in HADS 
compared to normal $\delta$~Sct stars can only be answered by nonlinear, 
nonradial pulsation models which do not yet exist. 

\section{Conclusions}

Using the full four-year light curves and periodograms of stars in the {\it
Kepler} field, all stars with {\it Kepler} magnitude $K_p < 12.5$\,mag, as 
well as many more fainter stars, were classified according to variability type.  
A catalog of $\delta$~Sct and $\gamma$~Dor stars with Gaia DR2 parallaxes is 
presented. From these data, luminosities, $\log L/L_\odot$, with a standard 
deviation of 0.04 dex.  By contrast, previous luminosity  estimates based on 
multicolour photometry have typical errors of 0.4 dex.

The most surprising result is that the $\gamma$~Dor variables do not occupy
a separate instability strip, but lie entirely within the $\delta$~Sct 
instability region.  There are, in fact, more $\delta$~Sct stars inside the 
$\gamma$~Dor instability region than $\gamma$~Dor stars.  No two classes of 
pulsating star are known to share the same instability region.  This must be
the case if the driving and damping mechanisms differ, as they do in the
conventional explanation for the two classes: the opacity $\kappa$~mechanism
for $\delta$~Sct stars and the convective blocking mechanism for
$\gamma$~Dor stars.

It seems that the frequencies in $\gamma$~Dor stars may be just an effect of 
mode selection rather than reflecting different driving and damping
mechanisms.  The presence of a large variety of light curves among the
$\gamma$~Dor stars, classified here as GDORA, GDORS and GDORM is probably an
indication of the sensitivity of mode selection to the conditions in the
outer layers of the star.  The GDORA type, for example, shows extreme
non-linear effects.  Why such nonlinearity exists only in a subset of these 
stars is not known.  This effect is not seen in $\delta$~Sct stars, but it 
might be masked by the presence of other frequencies of higher amplitudes.  
The suggestion by \citet{Xiong2016} that $\gamma$~Dor  stars should not be 
seen as a separate class has great merit.  Nevertheless, the distinction is 
still a useful one for classification purposes.  

Our understanding of stellar pulsation has evolved quite considerably over 
the last few decades.  It is now recognized that multiple driving and damping 
mechanisms occur in the same star.  The most important recent works in this 
respect are that of \citet{Xiong2015} and  \citet{Xiong2016} where it is 
demonstrated that the interplay of different processes can account for the 
instability at low frequencies seen in practically all $\delta$~Sct stars.  
In fact, inspection of the {\it Kepler} light curves shows that low
frequencies are present in at least 98\,percent of $\delta$~Sct stars.

The work of \citet{Xiong2016} offers a very attractive explanation for the 
low frequencies in $\delta$~Sct stars.  Unfortunately, the predicted 
red and blue edges from \citet{Xiong2016} do not agree with the limits of the
instability regions determined in this paper.  It is also not clear whether
an explanation for the co-existence of $\gamma$~Dor, $\delta$~Sct and
non-pulsating stars can be found by tuning available free parameters in the
theory.  Nevertheless, this work shows great promise for a better
understanding of these stars.

Another well-known group among the $\delta$~Sct stars are the HADS.  In
this paper it is shown that most HADS are normal $\delta$~Sct stars.

The data analyzed in this paper indicates that the majority of stars within
the $\delta$~Sct instability region do not pulsate.  If it is assumed that
these stars actually pulsate below the detectable level, then they should be
included in the calculation of the distribution of maximum amplitude.  In
that case, the very large number of apparently non-pulsating stars introduces 
a nonphysical discontinuity in the distribution of maximum amplitudes.  This
suggests that these stars do not pulsate.  This introduces yet another
unresolved issue because a reason needs to be found for the high damping of
pulsations in the majority of stars in the instability region.

It appears that one or more unknown damping and driving processes are 
operating in the outer layers, rendering the pulsation amplitudes and mode 
selection very sensitive to conditions in these layers.  These problems are 
perhaps not too surprising in view of the fact that starspots are present in 
most A stars \citep{Balona2017a}.  None of the current models of A stars 
provide a possible explanation for the presence of starspots.  The  problems 
discussed here add to the need for a revision in the current  view of A star 
atmospheres.

\section*{Acknowledgments}

LAB wishes to thank the National Research Foundation of South Africa for 
financial support.  This work has made use of data from the European Space 
Agency (ESA) mission Gaia (\url{https://www.cosmos.esa.int/gaia}), 
processed by the Gaia Data Processing and Analysis Consortium (DPAC,
\url{https://www.cosmos.esa.int/web/gaia/dpac/consortium}). Funding for the
DPAC has been provided by national institutions, in particular the institutions
participating in the Gaia Multilateral Agreement.  This paper includes 
data collected by the Kepler mission. Funding for the Kepler mission is 
provided by the NASA Science Mission directorate.

\bibliographystyle{mn2e}
\bibliography{pulaf}

\begin{thebibliography}{38}
\expandafter\ifx\csname natexlab\endcsname\relax\def\natexlab#1{#1}\fi

\bibitem[{{Bailer-Jones} {et~al.}(2013){Bailer-Jones}, {Andrae}, {Arcay},
  {Astraatmadja}, {Bellas-Velidis}, {Berihuete}, {Bijaoui}, {Carri{\'o}n},
  {Dafonte}, {Damerdji}, {Dapergolas}, {de Laverny}, {Delchambre}, {Drazinos},
  {Drimmel}, {Fr{\'e}mat}, {Fustes}, {Garc{\'{\i}}a-Torres}, {Gu{\'e}d{\'e}},
  {Heiter}, {Janotto}, {Karampelas}, {Kim}, {Knude}, {Kolka}, {Kontizas},
  {Kontizas}, {Korn}, {Lanzafame}, {Lebreton}, {Lindstr{\o}m}, {Liu},
  {Livanou}, {Lobel}, {Manteiga}, {Martayan}, {Ordenovic}, {Pichon},
  {Recio-Blanco}, {Rocca-Volmerange}, {Sarro}, {Smith}, {Sordo}, {Soubiran},
  {Surdej}, {Th{\'e}venin}, {Tsalmantza}, {Vallenari}, \&
  {Zorec}}]{Bailer-Jones2013}
{Bailer-Jones} C.~A.~L., {Andrae} R., {Arcay} B., {Astraatmadja} T.,
  {Bellas-Velidis} I., {Berihuete} A., {Bijaoui} A., {Carri{\'o}n} C.,
  {Dafonte} C., {Damerdji} Y., {Dapergolas} A., {de Laverny} P., {Delchambre}
  L., {Drazinos} P., {Drimmel} R., {Fr{\'e}mat} Y., {Fustes} D.,
  {Garc{\'{\i}}a-Torres} M., {Gu{\'e}d{\'e}} C., {Heiter} U., {Janotto} A.-M.,
  {Karampelas} A., {Kim} D.-W., {Knude} J., {Kolka} I., {Kontizas} E.,
  {Kontizas} M., {Korn} A.~J., {Lanzafame} A.~C., {Lebreton} Y., {Lindstr{\o}m}
  H., {Liu} C., {Livanou} E., {Lobel} A., {Manteiga} M., {Martayan} C.,
  {Ordenovic} C., {Pichon} B., {Recio-Blanco} A., {Rocca-Volmerange} B.,
  {Sarro} L.~M., {Smith} K., {Sordo} R., {Soubiran} C., {Surdej} J.,
  {Th{\'e}venin} F., {Tsalmantza} P., {Vallenari} A., {Zorec} J., 2013, \aap,
  559, A74

\bibitem[{{Balona}(2014)}]{Balona2014a}
{Balona} L.~A., 2014, \mnras, 437, 1476

\bibitem[{{Balona}(2017)}]{Balona2017a}
---, 2017, \mnras, 467, 1830

\bibitem[{{Balona}(2018)}]{Balona2018b}
---, 2018, \mnras, 476, 4840

\bibitem[{{Balona} \& {Abedigamba}(2016)}]{Balona2016b}
{Balona} L.~A., {Abedigamba} O.~P., 2016, \mnras, 461, 497

\bibitem[{{Balona} {et~al.}(2015){Balona}, {Daszy{\'n}ska-Daszkiewicz}, \&
  {Pamyatnykh}}]{Balona2015d}
{Balona} L.~A., {Daszy{\'n}ska-Daszkiewicz} J., {Pamyatnykh} A.~A., 2015,
  \mnras, 452, 3073

\bibitem[{{Balona} \& {Dziembowski}(2011)}]{Balona2011g}
{Balona} L.~A., {Dziembowski} W.~A., 2011, \mnras, 417, 591

\bibitem[{{Balona} {et~al.}(2016){Balona}, {Engelbrecht}, {Joshi}, {Joshi},
  {Sharma}, {Semenko}, {Pandey}, {Chakradhari}, {Mkrtichian}, {Hema}, \&
  {Nemec}}]{Balona2016c}
{Balona} L.~A., {Engelbrecht} C.~A., {Joshi} Y.~C., {Joshi} S., {Sharma} K.,
  {Semenko} E., {Pandey} G., {Chakradhari} N.~K., {Mkrtichian} D., {Hema}
  B.~P., {Nemec} J.~M., 2016, \mnras, 460, 1318

\bibitem[{{Balona} {et~al.}(2011){Balona}, {Guzik}, {Uytterhoeven}, {Smith},
  {Tenenbaum}, \& {Twicken}}]{Balona2011f}
{Balona} L.~A., {Guzik} J.~A., {Uytterhoeven} K., {Smith} J.~C., {Tenenbaum}
  P., {Twicken} J.~D., 2011, \mnras, 415, 3531

\bibitem[{{Bertelli} {et~al.}(2008){Bertelli}, {Girardi}, {Marigo}, \&
  {Nasi}}]{Bertelli2008}
{Bertelli} G., {Girardi} L., {Marigo} P., {Nasi} E., 2008, \aap, 484, 815

\bibitem[{{Brown} {et~al.}(2011){Brown}, {Latham}, {Everett}, \&
  {Esquerdo}}]{Brown2011a}
{Brown} T.~M., {Latham} D.~W., {Everett} M.~E., {Esquerdo} G.~A., 2011, \aj,
  142, 112

\bibitem[{{Castelli} \& {Kurucz}(2003)}]{Castelli2003}
{Castelli} F., {Kurucz} R.~L., 2003, in IAU Symposium, Vol. 210, Modelling of
  Stellar Atmospheres, {Piskunov} N., {Weiss} W.~W., {Gray} D.~F., eds., p. A20

\bibitem[{{Dupret} {et~al.}(2005{\natexlab{a}}){Dupret}, {Grigahc{\`e}ne},
  {Garrido}, {Gabriel}, \& {Scuflaire}}]{Dupret2005}
{Dupret} M., {Grigahc{\`e}ne} A., {Garrido} R., {Gabriel} M., {Scuflaire} R.,
  2005{\natexlab{a}}, \aap, 435, 927

\bibitem[{{Dupret} {et~al.}(2005{\natexlab{b}}){Dupret}, {Grigahc{\`e}ne},
  {Garrido}, {De Ridder}, {Scuflaire}, \& {Gabriel}}]{Dupret2005b}
{Dupret} M.-A., {Grigahc{\`e}ne} A., {Garrido} R., {De Ridder} J., {Scuflaire}
  R., {Gabriel} M., 2005{\natexlab{b}}, \mnras, 361, 476

\bibitem[{{Gaia Collaboration} {et~al.}(2018){Gaia Collaboration}, {Brown},
  {Vallenari}, {Prusti}, {de Bruijne}, {Babusiaux}, \&
  {Bailer-Jones}}]{Gaia2018}
{Gaia Collaboration}, {Brown} A.~G.~A., {Vallenari} A., {Prusti} T., {de
  Bruijne} J.~H.~J., {Babusiaux} C., {Bailer-Jones} C.~A.~L., 2018, ArXiv
  e-prints

\bibitem[{{Gaia Collaboration} {et~al.}(2016){Gaia Collaboration}, {Prusti},
  {de Bruijne}, {Brown}, {Vallenari}, {Babusiaux}, {Bailer-Jones}, {Bastian},
  {Biermann}, {Evans}, \& et~al.}]{Gaia2016}
{Gaia Collaboration}, {Prusti} T., {de Bruijne} J.~H.~J., {Brown} A.~G.~A.,
  {Vallenari} A., {Babusiaux} C., {Bailer-Jones} C.~A.~L., {Bastian} U.,
  {Biermann} M., {Evans} D.~W., et~al., 2016, \aap, 595, A1

\bibitem[{{Gontcharov}(2017)}]{Gontcharov2017}
{Gontcharov} G.~A., 2017, Astronomy Letters, 43, 472

\bibitem[{{Grigahc{\`e}ne} {et~al.}(2010){Grigahc{\`e}ne}, {Antoci}, {Balona},
  {Catanzaro}, {Daszy{\'n}ska-Daszkiewicz}, {Guzik}, {Handler}, {Houdek},
  {Kurtz}, {Marconi}, {Monteiro}, {Moya}, {Ripepi}, {Su{\'a}rez},
  {Uytterhoeven}, {Borucki}, {Brown}, {Christensen-Dalsgaard}, {Gilliland},
  {Jenkins}, {Kjeldsen}, {Koch}, {Bernabei}, {Bradley}, {Breger}, {Di
  Criscienzo}, {Dupret}, {Garc{\'{\i}}a}, {Garc{\'{\i}}a Hern{\'a}ndez},
  {Jackiewicz}, {Kaiser}, {Lehmann}, {Mart{\'{\i}}n-Ruiz}, {Mathias},
  {Molenda-{\.Z}akowicz}, {Nemec}, {Nuspl}, {Papar{\'o}}, {Roth}, {Szab{\'o}},
  {Suran}, \& {Ventura}}]{Grigahcene2010}
{Grigahc{\`e}ne} A., {Antoci} V., {Balona} L., {Catanzaro} G.,
  {Daszy{\'n}ska-Daszkiewicz} J., {Guzik} J.~A., {Handler} G., {Houdek} G.,
  {Kurtz} D.~W., {Marconi} M., {Monteiro} M.~J.~P.~F.~G., {Moya} A., {Ripepi}
  V., {Su{\'a}rez} J., {Uytterhoeven} K., {Borucki} W.~J., {Brown} T.~M.,
  {Christensen-Dalsgaard} J., {Gilliland} R.~L., {Jenkins} J.~M., {Kjeldsen}
  H., {Koch} D., {Bernabei} S., {Bradley} P., {Breger} M., {Di Criscienzo} M.,
  {Dupret} M., {Garc{\'{\i}}a} R.~A., {Garc{\'{\i}}a Hern{\'a}ndez} A.,
  {Jackiewicz} J., {Kaiser} A., {Lehmann} H., {Mart{\'{\i}}n-Ruiz} S.,
  {Mathias} P., {Molenda-{\.Z}akowicz} J., {Nemec} J.~M., {Nuspl} J.,
  {Papar{\'o}} M., {Roth} M., {Szab{\'o}} R., {Suran} M.~D., {Ventura} R.,
  2010, \apjl, 713, L192

\bibitem[{{Guzik} {et~al.}(2014){Guzik}, {Bradley}, {Jackiewicz},
  {Uytterhoeven}, \& {Kinemuchi}}]{Guzik2014}
{Guzik} J.~A., {Bradley} P.~A., {Jackiewicz} J., {Uytterhoeven} K., {Kinemuchi}
  K., 2014, in IAU Symposium, Vol. 301, Precision Asteroseismology, {Guzik}
  J.~A., {Chaplin} W.~J., {Handler} G., {Pigulski} A., eds., pp. 63--66

\bibitem[{{Guzik} {et~al.}(2000){Guzik}, {Kaye}, {Bradley}, {Cox}, \&
  {Neuforge}}]{Guzik2000}
{Guzik} J.~A., {Kaye} A.~B., {Bradley} P.~A., {Cox} A.~N., {Neuforge} C., 2000,
  \apjl, 542, L57

\bibitem[{{Handler} {et~al.}(2002){Handler}, {Balona}, {Shobbrook}, {Koen},
  {Bruch}, {Romero-Colmenero}, {Pamyatnykh}, {Willems}, {Eyer}, {James}, \&
  {Maas}}]{Handler2002b}
{Handler} G., {Balona} L.~A., {Shobbrook} R.~R., {Koen} C., {Bruch} A.,
  {Romero-Colmenero} E., {Pamyatnykh} A.~A., {Willems} B., {Eyer} L., {James}
  D.~J., {Maas} T., 2002, \mnras, 333, 262

\bibitem[{{Houdek}(2000)}]{Houdek2000}
{Houdek} G., 2000, in Astronomical Society of the Pacific Conference Series,
  Vol. 210, Delta Scuti and Related Stars, {M.~Breger \& M.~Montgomery}, ed.,
  pp. 454--+

\bibitem[{{Houdek} \& {Dupret}(2015)}]{Houdek2015}
{Houdek} G., {Dupret} M.-A., 2015, Living Reviews in Solar Physics, 12

\bibitem[{{Huber} {et~al.}(2014){Huber}, {Silva Aguirre}, {Matthews},
  {Pinsonneault}, {Gaidos}, {Garc{\'{\i}}a}, {Hekker}, {Mathur}, {Mosser},
  {Torres}, {Bastien}, {Basu}, {Bedding}, {Chaplin}, {Demory}, {Fleming},
  {Guo}, {Mann}, {Rowe}, {Serenelli}, {Smith}, \& {Stello}}]{Huber2014}
{Huber} D., {Silva Aguirre} V., {Matthews} J.~M., {Pinsonneault} M.~H.,
  {Gaidos} E., {Garc{\'{\i}}a} R.~A., {Hekker} S., {Mathur} S., {Mosser} B.,
  {Torres} G., {Bastien} F.~A., {Basu} S., {Bedding} T.~R., {Chaplin} W.~J.,
  {Demory} B.-O., {Fleming} S.~W., {Guo} Z., {Mann} A.~W., {Rowe} J.~F.,
  {Serenelli} A.~M., {Smith} M.~A., {Stello} D., 2014, \apjs, 211, 2

\bibitem[{{Mowlavi} {et~al.}(2013){Mowlavi}, {Barblan}, {Saesen}, \&
  {Eyer}}]{Mowlavi2013}
{Mowlavi} N., {Barblan} F., {Saesen} S., {Eyer} L., 2013, \aap, 554, A108

\bibitem[{{Murphy} {et~al.}(2015){Murphy}, {Bedding}, {Niemczura}, {Kurtz}, \&
  {Smalley}}]{Murphy2015}
{Murphy} S.~J., {Bedding} T.~R., {Niemczura} E., {Kurtz} D.~W., {Smalley} B.,
  2015, \mnras, 447, 3948

\bibitem[{{Papaloizou} \& {Pringle}(1978)}]{Papaloizou1978}
{Papaloizou} J., {Pringle} J.~E., 1978, \mnras, 182, 423

\bibitem[{{Pinsonneault} {et~al.}(2012){Pinsonneault}, {An},
  {Molenda-{\.Z}akowicz}, {Chaplin}, {Metcalfe}, \&
  {Bruntt}}]{Pinsonneault2012a}
{Pinsonneault} M.~H., {An} D., {Molenda-{\.Z}akowicz} J., {Chaplin} W.~J.,
  {Metcalfe} T.~S., {Bruntt} H., 2012, \apjs, 199, 30

\bibitem[{{Qian} {et~al.}(2018){Qian}, {Li}, {He}, {Zhang}, {Zhu}, \&
  {Han}}]{Qian2018}
{Qian} S.-B., {Li} L.-J., {He} J.-J., {Zhang} J., {Zhu} L.-Y., {Han} Z.-T.,
  2018, \mnras, 475, 478

\bibitem[{{Rodriguez} {et~al.}(2000){Rodriguez}, {Lopez-Gonzalez}, \& {Lopez de
  Coca}}]{Rodriguez2000}
{Rodriguez} E., {Lopez-Gonzalez} M.~J., {Lopez de Coca} P., 2000, VizieR Online
  Data Catalog, 414

\bibitem[{{Saio} {et~al.}(2018){Saio}, {Kurtz}, {Murphy}, {Antoci}, \&
  {Lee}}]{Saio2018a}
{Saio} H., {Kurtz} D.~W., {Murphy} S.~J., {Antoci} V.~L., {Lee} U., 2018,
  \mnras, 474, 2774

\bibitem[{{Smith} {et~al.}(2012){Smith}, {Stumpe}, {Van Cleve}, {Jenkins},
  {Barclay}, {Fanelli}, {Girouard}, {Kolodziejczak}, {McCauliff}, {Morris}, \&
  {Twicken}}]{Smith2012}
{Smith} J.~C., {Stumpe} M.~C., {Van Cleve} J.~E., {Jenkins} J.~M., {Barclay}
  T.~S., {Fanelli} M.~N., {Girouard} F.~R., {Kolodziejczak} J.~J., {McCauliff}
  S.~D., {Morris} R.~L., {Twicken} J.~D., 2012, \pasp, 124, 1000

\bibitem[{{Stumpe} {et~al.}(2012){Stumpe}, {Smith}, {Van Cleve}, {Twicken},
  {Barclay}, {Fanelli}, {Girouard}, {Jenkins}, {Kolodziejczak}, {McCauliff}, \&
  {Morris}}]{Stumpe2012}
{Stumpe} M.~C., {Smith} J.~C., {Van Cleve} J.~E., {Twicken} J.~D., {Barclay}
  T.~S., {Fanelli} M.~N., {Girouard} F.~R., {Jenkins} J.~M., {Kolodziejczak}
  J.~J., {McCauliff} S.~D., {Morris} R.~L., 2012, \pasp, 124, 985

\bibitem[{{Uytterhoeven} {et~al.}(2011){Uytterhoeven}, {Moya},
  {Grigahc{\`e}ne}, {Guzik}, {Guti{\'e}rrez-Soto}, {Smalley}, {Handler},
  {Balona}, {Niemczura}, {Fox Machado}, {Benatti}, {Chapellier}, {Tkachenko},
  {Szab{\'o}}, {Su{\'a}rez}, {Ripepi}, {Pascual}, {Mathias},
  {Mart{\'{\i}}n-Ru{\'{\i}}z}, {Lehmann}, {Jackiewicz}, {Hekker},
  {Gruberbauer}, {Garc{\'{\i}}a}, {Dumusque}, {D{\'{\i}}az-Fraile}, {Bradley},
  {Antoci}, {Roth}, {Leroy}, {Murphy}, {De Cat}, {Cuypers}, {Kjeldsen},
  {Christensen-Dalsgaard}, {Breger}, {Pigulski}, {Kiss}, {Still}, {Thompson},
  \& {van Cleve}}]{Uytterhoeven2011}
{Uytterhoeven} K., {Moya} A., {Grigahc{\`e}ne} A., {Guzik} J.~A.,
  {Guti{\'e}rrez-Soto} J., {Smalley} B., {Handler} G., {Balona} L.~A.,
  {Niemczura} E., {Fox Machado} L., {Benatti} S., {Chapellier} E., {Tkachenko}
  A., {Szab{\'o}} R., {Su{\'a}rez} J.~C., {Ripepi} V., {Pascual} J., {Mathias}
  P., {Mart{\'{\i}}n-Ru{\'{\i}}z} S., {Lehmann} H., {Jackiewicz} J., {Hekker}
  S., {Gruberbauer} M., {Garc{\'{\i}}a} R.~A., {Dumusque} X.,
  {D{\'{\i}}az-Fraile} D., {Bradley} P., {Antoci} V., {Roth} M., {Leroy} B.,
  {Murphy} S.~J., {De Cat} P., {Cuypers} J., {Kjeldsen} H.,
  {Christensen-Dalsgaard} J., {Breger} M., {Pigulski} A., {Kiss} L.~L., {Still}
  M., {Thompson} S.~E., {van Cleve} J., 2011, \aap, 534, A125

\bibitem[{{Xiong}(1989)}]{Xiong1989}
{Xiong} D.-R., 1989, \aap, 209, 126

\bibitem[{{Xiong} {et~al.}(1998){Xiong}, {Cheng}, \& {Deng}}]{Xiong1998b}
{Xiong} D.~R., {Cheng} Q.~L., {Deng} L., 1998, \apj, 500, 449

\bibitem[{{Xiong} {et~al.}(2015){Xiong}, {Deng}, \& {Zhang}}]{Xiong2015}
{Xiong} D.~R., {Deng} L., {Zhang} C., 2015, \mnras, 451, 3354

\bibitem[{{Xiong} {et~al.}(2016){Xiong}, {Deng}, {Zhang}, \&
  {Wang}}]{Xiong2016}
{Xiong} D.~R., {Deng} L., {Zhang} C., {Wang} K., 2016, \mnras, 457, 3163

\end{thebibliography}

\label{lastpage}

\end{document}